\documentclass[pdflatex,sn-mathphys-num]{sn-jnl}

\usepackage{graphicx}
\usepackage{multirow}
\usepackage{amsmath,amssymb,amsfonts}
\usepackage{amsthm}
\usepackage{mathrsfs}
\usepackage[title]{appendix}
\usepackage{xcolor}
\usepackage{textcomp}
\usepackage{manyfoot}
\usepackage{booktabs}
\usepackage{xspace}

\begin{document}

\title[Strange quark matter from a flavor-dressed contact interaction]{%
  Deconfinement Confined to the Heaviest Neutron Stars: Strange Quark
  Matter from a Flavor-Dressed Contact Interaction}

\author[1]{\fnm{Aftab} \sur{Ahmad}}\email{aftabahmad@gu.edu.pk}
\author[1]{\fnm{Noreen} \sur{Akhter}}\email{noreenakhtar3698@gmail.com}
\author[2,3]{\fnm{Alfredo} \sur{Raya}}\email{alfredo.raya@umich.mx}

\affil[1]{\orgdiv{Institute of Physics}, \orgname{Gomal University},
  \orgaddress{\city{D.I.\ Khan}, \postcode{29220},
  \state{Khyber Pakhtunkhwa}, \country{Pakistan}}}
\affil[2]{Facultad de Ingenier\'ia El\'ectrica,
  Universidad Michoacana de San Nicol\'as de Hidalgo,
  Edificio Omega, Ciudad Universitaria,
  Francisco J.\ M\'ujica s/n, Col.\ Fel\'icitas del R\'io,
  CP 58040 Morelia, Michoac\'an, M\'exico}
\affil[3]{Centro de Ciencias Exactas,
  Universidad del B\'io-B\'io, Casilla 447, Chill\'an, Chile}

\abstract{%
We study cold, electrically neutral, $\beta$-equilibrated quark matter
with $u$, $d$, and $s$ quarks described by a symmetry-preserving contact
interaction whose parameters are fixed in vacuum. The vector repulsion
is varied as a fraction $0 \le \eta \le 3/2$ of the scalar coupling,
and results are given with and without the flavor dressing of the
effective coupling. Pure quark matter is not self-bound in this model,
and no value of $\eta$ produces a star that reaches two solar masses
while satisfying the tidal constraint from GW170817; the heaviest
configurations owe their mass to an envelope of constituent quarks. We
then construct hybrid stars with a relativistic mean-field hadronic
phase fitted to GM1 saturation properties, using both Maxwell and Gibbs
constructions and a deconfinement bag shift. Tying deconfinement to
chiral restoration and requiring a continuous pressure yields an upper
bound on this shift. Within that bound, quark matter appears inside
stable stars only for $\eta \lesssim 1/2$, never below about $4\,n_0$,
and only in stars between 2.2 and $2.35\,M_\odot$, over a stable mass
window of at most $0.02\,M_\odot$. The vector strength that pure quark
stars need to become massive prevents deconfinement in hybrid stars.
Radii and tidal deformabilities of intermediate-mass stars are therefore
set by the hadronic EoS alone; with the softer SLy4 EoS the transition
moves to the maximum-mass configuration itself.}


\maketitle

%% ====================================================================
\section{Introduction}
\label{sec:intro}
%% ====================================================================

The equation of state (EoS) of strongly interacting matter at several
times the nuclear saturation density $n_0 \simeq 0.16\,\text{fm}^{-3}$
remains one of the central open problems of nuclear and particle physics.
It controls the structure of neutron stars, the dynamics of core-collapse
supernovae, and the waveform of gravitational radiation emitted by
merging compact binaries~\cite{Lattimer:2000nx,Ozel:2016oaf}. At these
densities perturbative QCD does not apply, and lattice QCD, which
establishes a chiral crossover at $T_{pc} \approx 155$\,MeV for vanishing
baryon chemical potential~\cite{Borsanyi:2013bia,HotQCD:2014kol}, runs
into the fermion sign problem as soon as the real baryon chemical
potential becomes large~\cite{Nagata:2021ugx}. Cold and dense matter is
therefore described with effective models, and the observable properties
of neutron stars provide the most direct test of those descriptions.

The observational situation has changed considerably over the last
decade. Pulsar timing has established masses close to or above  ($2M_\odot$) for PSR~J1614$-$2230~\cite{Demorest:2010bx,Arzoumanian:2017puf},
PSR~J0348$+$0432~\cite{Antoniadis:2013pzd}, and
PSR~J0740$+$6620, whose mass is $2.08 \pm 0.07\,M_\odot$~\cite{Fonseca:2021wxt}.
The NICER mission has added joint mass and radius estimates for
PSR~J0030$+$0451 and PSR~J0740$+$6620~\cite{Riley:2019yda,Miller:2019cac,Riley:2021pdl},
with central radius estimates between about 12.4 and 13.7\,km.
Analyses of the larger NICER data set now available for
PSR~J0740$+$6620 give $12.49^{+1.28}_{-0.88}$\,km~\cite{Salmi:2024aum}
and $12.92^{+2.09}_{-1.13}$\,km~\cite{Dittmann:2024mbo}, consistent with
the earlier estimates. The
gravitational-wave event GW170817 constrains the tidal deformability of
a $1.4\,M_\odot$ star to $\Lambda_{1.4} = 190^{+390}_{-120}$ at 90\%
credibility~\cite{LIGOScientific:2018cki}. Taken together, these data
demand an EoS that is stiff enough to support $2M_\odot$ yet soft
enough at intermediate densities to keep $\Lambda_{1.4}$ below about 580.
Whether quark matter plays any role in meeting both requirements is
still an open question~\cite{Baym:2017whm,Annala:2019puf}.

Continuum approaches to QCD offer a way to model quark matter while
keeping contact with the dynamics of dynamical chiral symmetry breaking.
Within the framework of Schwinger--Dyson equations~\cite{Roberts:1994dr,Fischer:2018sdj},
the contact interaction (CI) replaces the momentum-dependent gluon
exchange by a symmetry-preserving, ultra-local kernel regularized with a
proper-time scheme. Despite its simplicity, the CI reproduces a broad
set of light and strange hadron properties once its infrared coupling
and regulators are fixed in vacuum~\cite{GutierrezGuerrero:2010md,Roberts:2010rn,Roberts:2011cf,Chen:2012qr},
and it has been applied to the QCD phase diagram at finite temperature,
chemical potential, and external fields~\cite{Ahmad:2016iez,Ahmad:2020ifp,Ahmad:2023ecw,ahmad2025qcdphasediagramlarge,Ahmad:2025magn}.
%\cite{Ahmad:2016iez,Ahmad:2020ifp,Ahmad:2023ecw,Ahmad}
A refinement relevant for the present work is the flavor-dressed CI, in
which the effective coupling decreases with the number of active flavors
$N_f$ so as to reproduce the suppression of chiral symmetry breaking
expected near the conformal window~\cite{Ahmad:2020jzn,Ahmad:2024emu}.
For two light flavors this dressing reduces to the identity. With a
strange quark present it does not, and its effect on dense matter can
be quantified directly.

After the incorporation of the $s$-quark into the game, two questions follow 
naturally. The first is whether strange quark
matter described by the CI can form self-bound stars or support the
observed maximum masses on its own. The Bodmer--Witten hypothesis
allows, in principle, for three-flavor quark matter to be the absolute
ground state of strong interactions~\cite{Bodmer:1971we,Witten:1984rs},
and effective models with vector repulsion have been used to build
$2M_\odot$ quark stars~\cite{Klahn:2015mfa,universe4020030}. The
second question concerns hybrid stars, in which a quark core forms
inside a hadronic envelope. Here the location of the deconfinement
transition relative to the central density of the heaviest stable
configuration decides everything, and models of the Nambu--Jona-Lasinio
type have often found that quark matter either appears only near the
maximum mass or destabilizes the star as soon as it
appears~\cite{Buballa:2003qv,Alford:2004pf,Pagliara:2007ph,Baym:2017whm}.

In this paper we address both questions with a CI that includes the
$u$, $d$, and $s$ quarks, repulsive vector interactions of strength
$G_V = \eta\, G_S$, electric charge neutrality, and $\beta$ equilibrium
with electrons and muons. All vacuum parameters are fixed beforehand;
with a current strange mass $m_s = 170$\,MeV the model yields a
dressed strange mass $M_s = 533$\,MeV, in line with earlier CI studies of
strange hadrons~\cite{Chen:2012qr}. We scan $\eta$ between 0 and $3/2$
and compare the results with and without flavor dressing. For hybrid
stars we describe the hadronic phase with a relativistic mean-field
model fitted to the GM1 saturation properties~\cite{Glendenning:1991es},
join it to the quark phase through both a Maxwell and a Gibbs
construction~\cite{Glendenning:1992vb}, and allow for a deconfinement
bag shift $B_\text{dc}$ in the spirit of the vBag
model~\cite{Klahn:2015mfa}. Requiring that the pressure remain
continuous in the baryon chemical potential sets an upper bound on
$B_\text{dc}$ for each value of $\eta$, which removes most of the
freedom that such a shift would otherwise introduce.

The outcome is unambiguous within this setup. Pure CI quark matter is
not self-bound, and no value of $\eta$ reconciles $2M_\odot$ with
the tidal constraint from GW170817. When a hadronic envelope is added,
the quark phase appears inside stable stars only for
$\eta \lesssim 1/2$, and never below about $4\,n_0$. Stable quark cores then exist in stars between 2.2
and $2.35\,M_\odot$, over a mass window never wider than a few
hundredths of a solar mass. Stars of intermediate mass remain purely
hadronic, so their radii and tidal deformabilities carry no imprint of
the quark phase. The vector strength that pure quark stars need to reach
$2M_\odot$ is precisely the one that prevents deconfinement inside
stable hybrid stars.
For the presentetion of ideas, we organized the remaining of the paper 
 as follows. Section~\ref{sec:model} presents the
flavor-dressed CI with three flavors and its thermodynamic potential.
Section~\ref{sec:quarkmatter} constructs neutral quark matter in $\beta$
equilibrium, and Section~\ref{sec:quarkstars} discusses pure quark stars.
The hadronic model and the phase transition are introduced in
Section~\ref{sec:transition}, while Section~\ref{sec:hybrid} presents
hybrid stars and confronts them with observations. Our conclusions are
drawn in Section~\ref{sec:conclusions}, and the numerical checks are
collected in Appendix~\ref{app:checks}.

%% ====================================================================
\section{Flavor-dressed contact interaction with three flavors}
\label{sec:model}
%% ====================================================================

\subsection{Gap equation and flavor dressing}
\label{subsec:gap}

In Euclidean space the dressed propagator of a quark of flavor $f$
obeys the Schwinger--Dyson equation
\begin{equation}
S_f^{-1}(p) = i\gamma\cdot p + m_f
+ \frac{4}{3}\int\!\frac{d^4q}{(2\pi)^4}\,
g^2 D_{\mu\nu}(p-q)\,\gamma_\mu S_f(q)\,\Gamma_\nu(q,p),
\label{eq:SDE}
\end{equation}
where $m_f$ is the current mass, $4/3$ is the color factor for
$N_c=3$, $D_{\mu\nu}$ is the dressed gluon propagator, and
$\Gamma_\nu$ is the dressed quark-gluon vertex~\cite{Roberts:1994dr,Fischer:2018sdj}.
The contact interaction keeps the rainbow vertex, $\Gamma_\nu \to \gamma_\nu$,
and replaces the gluon exchange by its infrared limit, in which the
gluon carries a dynamically generated mass
$m_g$~\cite{GutierrezGuerrero:2010md,Roberts:2011cf,Aguilar:2010gm},
\begin{equation}
g^2 D_{\mu\nu}(k) \;\to\; \delta_{\mu\nu}\,\frac{4\pi\alpha_{\rm ir}}{m_g^2}
\equiv \delta_{\mu\nu}\,\alpha_{\rm eff}.
\label{eq:kernel}
\end{equation}
We use $\alpha_{\rm ir}=0.93\pi$ and $m_g = 0.8$\,GeV, the values fixed
by vacuum meson phenomenology~\cite{Roberts:2011cf,Chen:2012qr},
which give $\alpha_{\rm eff}=57.4\,\text{GeV}^{-2}$. With a
momentum-independent kernel the propagator takes the form
$S_f^{-1} = i\gamma\cdot p + M_f$, and the dressed mass $M_f$ is a
constant determined by
\begin{equation}
M_f = m_f + \frac{16\,\alpha_{\rm eff}}{3}
\int\!\frac{d^4q}{(2\pi)^4}\,\frac{M_f}{q^2+M_f^2}.
\label{eq:gapvac}
\end{equation}
The effect of additional light flavors on dynamical chiral symmetry
breaking is modeled through a flavor-dependent coupling. Following
Refs.~\cite{Ahmad:2020jzn,Ahmad:2024emu,ahmad2025qcdphasediagramlarge} we write
\begin{equation}
\alpha_{\rm eff}(N_f) = \alpha_{\rm eff}
\sqrt{1-\frac{N_f-2}{\mathcal{N}_f^c}},
\qquad \mathcal{N}_f^c = N_f^c + \varsigma,
\label{eq:flavordress}
\end{equation}
with $\varsigma = 2.9$ chosen so that chiral symmetry is restored at
$N_f^c \approx 8$ in the chiral limit. The coupling equals
$\alpha_{\rm eff}$ for $N_f=2$. For the three-flavor matter studied here
it is reduced by a factor $0.953$. Since the vacuum parameters were
calibrated with two flavors, this reduction changes the vacuum masses,
and we therefore present every result both with the dressing
($N_f=3$ in Eq.~\eqref{eq:flavordress}) and without it
($\alpha_{\rm eff}$ unchanged). The two sets of results bracket the
uncertainty associated with this choice.

\subsection{Regularization and vacuum}
\label{subsec:reg}

The integral in Eq.~\eqref{eq:gapvac} is regularized with the
proper-time scheme of Refs.~\cite{GutierrezGuerrero:2010md,Roberts:2011cf},
\begin{equation}
\frac{1}{q^2+M^2} = \int_0^\infty\! d\tau\, e^{-\tau(q^2+M^2)}
\;\to\; \int_{\tau_{\rm uv}^2}^{\tau_{\rm ir}^2}\! d\tau\,
e^{-\tau(q^2+M^2)},
\label{eq:PT}
\end{equation}
with $\tau_{\rm uv} = 1/\Lambda_{\rm uv}$, $\Lambda_{\rm uv}=0.905$\,GeV,
and $\tau_{\rm ir} = 1/\Lambda_{\rm ir}$, $\Lambda_{\rm ir}=0.240$\,GeV.
The infrared cutoff removes quark production thresholds and so
implements a simple form of confinement (see, for example, a recent  comparative studies on confinement from various model of low energy QCD in ~\cite{Ahmad:2026puc}), while the ultraviolet cutoff
sets the hadronic scale of the model. The momentum integration can then
be done in closed form, and Eq.~\eqref{eq:gapvac} becomes
\begin{equation}
M_f = m_f + \frac{\alpha_{\rm eff}}{3\pi^2}\,M_f\,
\mathcal{C}(M_f^2),
\label{eq:gapPT}
\end{equation}
where
\begin{equation}
\mathcal{C}(M^2) = \int_{\tau_{\rm uv}^2}^{\tau_{\rm ir}^2}
\frac{d\tau}{\tau^2}\,e^{-\tau M^2}
= \frac{e^{-M^2\tau_{\rm uv}^2}}{\tau_{\rm uv}^2}
-\frac{e^{-M^2\tau_{\rm ir}^2}}{\tau_{\rm ir}^2}
+M^2\left[\mathrm{Ei}\!\left(-M^2\tau_{\rm uv}^2\right)
-\mathrm{Ei}\!\left(-M^2\tau_{\rm ir}^2\right)\right],
\label{eq:Cfun}
\end{equation}
and $\mathrm{Ei}(x)$ is the exponential integral. For the vacuum energy we
also need
\begin{equation}
\mathcal{D}(M^2) = \int_{\tau_{\rm uv}^2}^{\tau_{\rm ir}^2}
\frac{d\tau}{\tau^3}\,e^{-\tau M^2},
\qquad \frac{d\mathcal{D}}{dM^2} = -\,\mathcal{C}(M^2),
\label{eq:Dfun}
\end{equation}
which also has a closed form in terms of exponentials and $\mathrm{Ei}(x)$.
With $m_u=m_d=7$\,MeV the vacuum solution is $M_u=M_d=367.5$\,MeV, and
the condensate $-\langle\bar q q\rangle = 9(M_u-m_u)/(4\alpha_{\rm eff})$
gives $(-\langle\bar q q\rangle)^{1/3}=242$\,MeV. For the strange quark we
take $m_s=170$\,MeV, the value used in CI studies of strange
hadrons~\cite{Chen:2012qr}, and obtain $M_s=533$\,MeV. With flavor
dressing these masses become $M_u = 343.2$\,MeV and $M_s = 518.5$\,MeV.
Table~\ref{tab:vacuum} collects the vacuum quantities.

\begin{table}[t]
\caption{Vacuum properties of the contact interaction with and without
flavor dressing for $N_f=3$. Masses in MeV, coupling in GeV$^{-2}$.}
\label{tab:vacuum}
\begin{tabular*}{\textwidth}{@{\extracolsep\fill}lcccc}
\toprule
 & $\alpha_{\rm eff}$ & $M_u=M_d$ & $M_s$ & $(-\langle\bar q q\rangle)^{1/3}$ \\
\midrule
Without dressing & 57.37 & 367.5 & 533.0 & 242 \\
With dressing ($N_f=3$) & 54.67 & 343.2 & 518.5 & 240 \\
\botrule
\end{tabular*}
\end{table}

\subsection{Thermodynamic potential at finite density}
\label{subsec:omega}

After a Fierz rearrangement of the contact vertex, the rainbow gap
equation coincides with the mean-field gap equation of a four-fermion
model with scalar coupling $G_S = 2\alpha_{\rm eff}/9$ in each flavor
channel~\cite{Buballa:2003qv,Klahn:2015mfa}. The same rearrangement
produces a repulsive vector coupling $G_V = \alpha_{\rm eff}/9$, that is,
$G_V/G_S = 1/2$. Because the vector channel of the full QCD kernel is
not fixed by the vacuum phenomenology that calibrates the CI, we treat
its strength as a parameter and write $G_V = \eta\, G_S$, with $\eta=1/2$
as the reference value. The rainbow kernel does not mix flavors, so at
zero temperature the thermodynamic potential is a sum of independent
flavor contributions,
$\Omega = \sum_{f} \Omega_f$, with
\begin{equation}
\Omega_f = \frac{(M_f-m_f)^2}{4G_S}
-\frac{(\mu_f-\tilde\mu_f)^2}{4G_V}
+\frac{N_c}{8\pi^2}\,\mathcal{D}(M_f^2)
-P_{\rm FG}(\tilde\mu_f,M_f).
\label{eq:omega}
\end{equation}
Here $\mu_f$ is the chemical potential of flavor $f$, $\tilde\mu_f$ the
chemical potential shifted by the vector mean field, and
\begin{equation}
P_{\rm FG}(\mu,M) = \frac{N_c}{24\pi^2}\left[\mu\, k_F
\left(2\mu^2-5M^2\right) + 3M^4 \ln\frac{\mu+k_F}{M}\right]
\theta(\mu-M),
\label{eq:PFG}
\end{equation}
with $k_F = \sqrt{\mu^2-M^2}$, is the pressure of a degenerate gas of
quasiparticles of mass $M$ with spin and color degeneracy $2N_c$. The
medium contribution is ultraviolet finite and is left unregularized, as
is customary. The stationarity conditions
$\partial\Omega_f/\partial M_f = \partial\Omega_f/\partial\tilde\mu_f = 0$
give
\begin{align}
M_f &= m_f + \frac{\alpha_{\rm eff}}{3\pi^2}\,M_f\,\mathcal{C}(M_f^2)
- 2G_S\, n_s(\tilde\mu_f,M_f),
\label{eq:gapmed} \\
\tilde\mu_f &= \mu_f - 2G_V\, n_f(\tilde\mu_f,M_f),
\label{eq:mutilde}
\end{align}
where
\begin{equation}
n_s = \frac{N_c M}{2\pi^2}\left[\mu\,k_F - M^2\ln\frac{\mu+k_F}{M}\right],
\qquad
n_f = \frac{N_c\,k_F^3}{3\pi^2},
\label{eq:densities}
\end{equation}
both evaluated at $\tilde\mu_f$. Equation~\eqref{eq:gapmed} reduces to
Eq.~\eqref{eq:gapPT} in vacuum, and the medium term, through Pauli
blocking, is what drives the restoration of chiral symmetry.
Equation~\eqref{eq:mutilde} expresses the vector repulsion: at fixed
$\mu_f$ it lowers the Fermi momentum and therefore the density.

When Eqs.~\eqref{eq:gapmed} and \eqref{eq:mutilde} admit several
solutions, the physical one is the solution with the lowest $\Omega_f$.
The pressure is measured with respect to the physical vacuum,
\begin{equation}
P_q(\{\mu_f\}) = -\sum_f\left[\Omega_f(\mu_f) - \Omega_f^{\rm vac}\right],
\label{eq:Pq}
\end{equation}
and the quark number densities follow from $n_f = \partial P_q/\partial\mu_f$.
We checked numerically that the density given by
Eq.~\eqref{eq:densities} equals the derivative of Eq.~\eqref{eq:Pq}
to a relative accuracy of $10^{-5}$ across the whole range of chemical
potentials used below. This check matters. With an incorrect relative
normalization between the condensate term and the vacuum energy in
Eq.~\eqref{eq:omega}, the gap equation can still be satisfied while the
pressure is off by a constant factor, and the resulting stellar masses
and radii are then meaningless.

\subsection{Chiral restoration at finite density}
\label{subsec:chiral}

\begin{figure}[t]
\centering
\includegraphics[width=\textwidth]{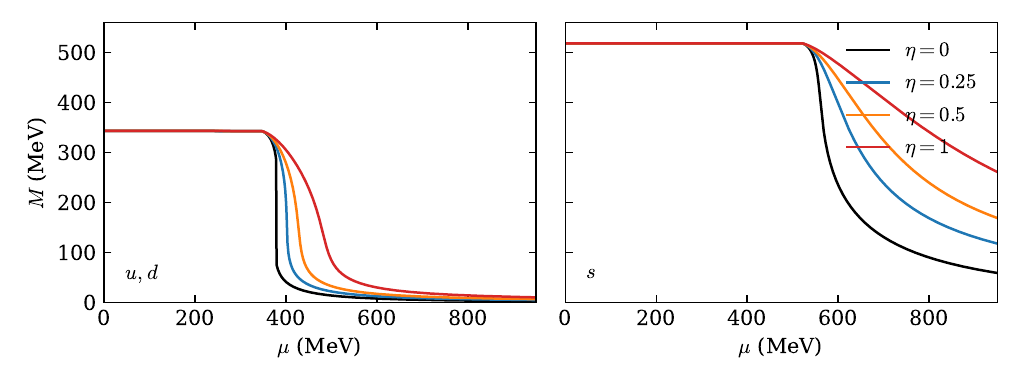}
\caption{Dressed masses of the light (left) and strange (right) quarks
as functions of the quark chemical potential, with flavor dressing, for
several values of $\eta = G_V/G_S$.}
\label{fig:masses}
\end{figure}

Figure~\ref{fig:masses} shows the dressed masses as functions of the
quark chemical potential for several values of $\eta$. Without vector
repulsion the light-quark transition is of first order, with a jump in
$M_u$ at $\mu = 397$\,MeV, or at 378\,MeV with flavor dressing. Vector
repulsion softens the transition and moves it to larger $\mu$. At
$\eta = 1/4$ it is already a continuous crossover in both cases,
although a steep one, particularly without dressing, where $M_u$ drops
by about 180\,MeV per MeV of chemical potential at its steepest point. We locate the light-quark transition at the value $\mu_\chi$
where $-dM_u/d\mu$ is largest; with flavor dressing we find
$\mu_\chi = 401$, 427, 481, and 535\,MeV for $\eta = 1/4$, $1/2$, 1, and
$3/2$. Above $\mu_\chi$ the light quarks carry a mass of a few tens of
MeV that decreases slowly towards $m_u$.

The strange quark behaves differently. Its larger current mass keeps
$M_s$ close to its vacuum value until $\mu \approx 560$--$700$\,MeV,
depending on $\eta$, and even beyond that point it stays heavy:
at $\mu = 950$\,MeV we find $M_s \approx 60$\,MeV for $\eta=0$ and
between 120 and 330\,MeV for $1/4 \le \eta \le 3/2$. In neutral matter this
delays the appearance of strangeness, as discussed in the next section.

%% ====================================================================
\section{Neutral quark matter in beta equilibrium}
\label{sec:quarkmatter}
%% ====================================================================

Matter in a neutron star is electrically neutral and in chemical
equilibrium under weak interactions. For $u$, $d$, and $s$ quarks
together with electrons and muons, $\beta$ equilibrium fixes
\begin{equation}
\mu_d = \mu_s = \mu_u + \mu_e, \qquad \mu_\mu = \mu_e,
\label{eq:beta}
\end{equation}
so that all chemical potentials are determined by the baryon chemical
potential $\mu_B = \mu_u + 2\mu_d$ and the electron chemical potential
$\mu_e$. The latter is fixed by charge neutrality,
\begin{equation}
\frac{2}{3}\,n_u - \frac{1}{3}\left(n_d + n_s\right) - n_e - n_\mu = 0.
\label{eq:neutrality}
\end{equation}
The leptons are treated as free Fermi gases. The total pressure is
$P = P_q + P_\ell$, with $P_q$ from Eq.~\eqref{eq:Pq}, and the energy
density and baryon density follow from
\begin{equation}
\varepsilon = -P + \sum_{i} \mu_i n_i, \qquad
n_B = \frac{1}{3}\left(n_u+n_d+n_s\right),
\label{eq:eps}
\end{equation}
where the sum runs over quarks and leptons. Because the rainbow kernel
does not couple different flavors, the flavors interact only through
Eqs.~\eqref{eq:beta} and \eqref{eq:neutrality}. Each quark flavor can
therefore be solved on a grid of $\mu_f$ once, and neutral matter is
then assembled by root finding in $\mu_e$ at every $\mu_B$.

The first thing to notice is where the pressure becomes positive. In
the chirally broken phase the quarks behave as quasiparticles with the
vacuum masses of Table~\ref{tab:vacuum}, and a Fermi sea forms as soon
as $\mu_u$ and $\mu_d$ exceed $M_u$. The pressure therefore turns
positive at $\mu_B \simeq 3M_u$, that is, at 1102\,MeV without dressing
and 1030\,MeV with it, independently of $\eta$. The energy per baryon
never falls below 1110 and 1037\,MeV, respectively. Both values lie well
above the 930\,MeV of the most bound nuclei, so CI strange quark matter
is not self-bound and the Bodmer--Witten
hypothesis~\cite{Bodmer:1971we,Witten:1984rs} is not realized in this
model. The result has a direct consequence for stellar structure. The
low-density phase of the model is a dilute gas of constituent quarks,
which has no counterpart in nature, and any stellar configuration whose
outer layers are made of it should not be taken at face value. We come
back to this point in Section~\ref{sec:quarkstars}.

Chiral symmetry is restored for the light quarks when $\mu_u$ and
$\mu_d$ cross the value $\mu_\chi$ discussed in
Section~\ref{subsec:chiral}. Strangeness appears later, once $\mu_d$
exceeds the in-medium strange mass. The corresponding baryon density
depends strongly on the vector coupling. With flavor dressing, strange
quarks appear at $n_B = 4.0\,n_0$ for $\eta = 1/4$, $3.2\,n_0$ for
$\eta=1/2$, $2.3\,n_0$ for $\eta=1$, and $1.6\,n_0$ for $\eta=3/2$;
without dressing the numbers differ by less than 10\%. The trend has a
simple origin. Vector repulsion raises the chemical potential needed to
reach a given density, and the strange threshold, which is fixed in
terms of $\mu_d$, is then crossed at a lower density. Once present, the
strange quarks reduce the need for electrons. The electron chemical
potential never exceeds 120\,MeV, so muons appear only in a narrow range
and in small numbers, with no visible effect on the EoS.

\begin{figure}[t]
\centering
\includegraphics[width=0.6\textwidth]{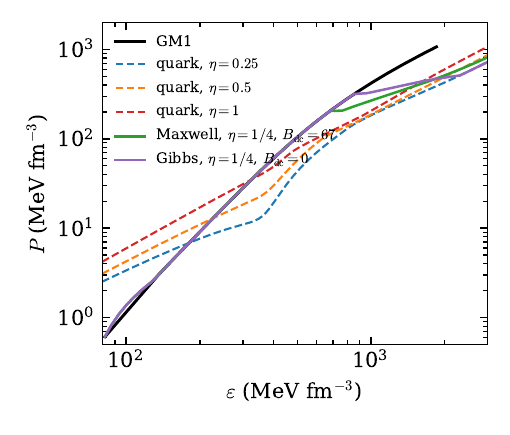}
\caption{Pressure versus energy density. Dashed lines: neutral quark
matter in $\beta$ equilibrium with flavor dressing for three values of
$\eta$. Solid black line: GM1 hadronic matter. Colored solid lines:
hybrid EoS with a Maxwell construction ($B_{\rm dc}=0.98\,B_{\max}$)
and with a Gibbs construction ($B_{\rm dc}=0$), both for
$\eta = 1/4$; see Section~\ref{sec:transition}.}
\label{fig:eos}
\end{figure}

Figure~\ref{fig:eos} shows the resulting pressure as a function of the
energy density. The vector interaction stiffens the EoS at high
density: for $\eta \ge 1/4$ the squared speed of sound exceeds the
conformal value $1/3$ and reaches about 0.5 at the highest densities
considered, while staying well below the causal limit. The same
interaction lowers the pressure at a given chemical potential, which
will matter for the competition with hadronic matter in
Section~\ref{sec:transition}.

The case $\eta = 0$ deserves a separate remark. The light-quark
transition is then of first order, and within the coexistence region no
homogeneous neutral solution exists; neutral matter would have to be
built as a mixed phase with two conserved charges. Since this limit
gives the softest EoS of the set, with $c_s^2$ never above $1/3$, and
since vector repulsion is generated by the Fierz rearrangement of the
kernel itself,
we keep $\eta = 0$ only as a reference for the EoS and do not compute
stellar configurations for it.

%% ====================================================================
\section{Pure quark stars}
\label{sec:quarkstars}
%% ====================================================================

\subsection{Stellar structure and tidal deformability}
\label{subsec:TOV}

For a static, spherically symmetric star the metric functions and the
pressure follow from the Tolman--Oppenheimer--Volkoff
equations~\cite{Tolman:1939jz,Oppenheimer:1939ne}, which in units
$G=c=1$ read
\begin{equation}
\frac{dP}{dr} = -\frac{\left[\varepsilon(P)+P\right]\left[m+4\pi r^3P\right]}
{r\left(r-2m\right)},
\qquad
\frac{dm}{dr} = 4\pi r^2\varepsilon(P).
\label{eq:TOV}
\end{equation}
The radius $R$ is the point where the pressure vanishes, and
$M = m(R)$. The response of the star to a static quadrupolar tidal field
is measured by the dimensionless tidal deformability
$\Lambda = (2/3)\,k_2\, C^{-5}$, where $C = M/R$ is the compactness and
$k_2$ the second Love number~\cite{Hinderer:2007mb}. We obtain $k_2$
from the function $y(r)$, which satisfies
\begin{equation}
r\frac{dy}{dr} + y^2 + y\,e^{\lambda}\left[1+4\pi r^2(P-\varepsilon)\right]
+ r^2 Q = 0,
\label{eq:yeq}
\end{equation}
with $y(0)=2$, $e^{\lambda} = (1-2m/r)^{-1}$, and
\begin{equation}
Q = 4\pi e^{\lambda}\left[5\varepsilon + 9P
+ \frac{\varepsilon+P}{c_s^2}\right]
- \frac{6\,e^{\lambda}}{r^2}
- \left[\frac{2 e^{\lambda}\left(m+4\pi r^3 P\right)}{r^2}\right]^2 .
\label{eq:Qfun}
\end{equation}
The Love number then follows from the standard expression in terms of
$C$ and $y_R = y(R)$~\cite{Hinderer:2007mb}. Wherever the energy density
jumps by $\Delta\varepsilon$ at fixed pressure, as happens at a Maxwell
phase transition or at the surface of a self-bound object, $1/c_s^2$
contains a delta function and Eq.~\eqref{eq:yeq} must be supplemented
by the matching condition
$y \to y - 4\pi r^3 \Delta\varepsilon / m$ at that
radius~\cite{Damour:2009vw,Postnikov:2010yn}. We integrate
Eqs.~\eqref{eq:TOV} and \eqref{eq:yeq} together and apply this
condition explicitly at every discontinuity. The numerical scheme
reproduces the known results for a bag-model EoS and for an $n=1$
polytrope in the Newtonian limit, as shown in Appendix~\ref{app:checks}.

\subsection{Mass--radius relation and tidal deformability}
\label{subsec:QSresults}

\begin{figure*}[t]
\centering
\includegraphics[width=\textwidth]{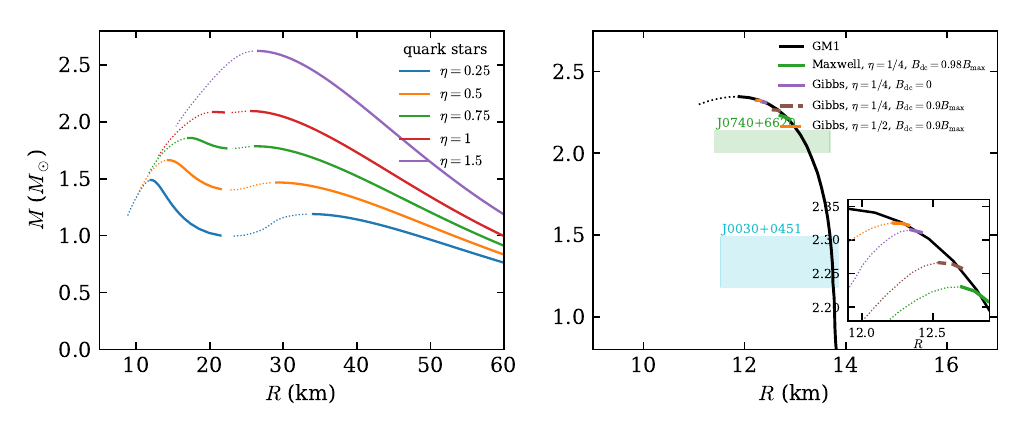}
\caption{Mass--radius relations. Left: pure quark stars with flavor
dressing; solid lines are stable configurations and dotted lines
unstable ones. Right: GM1 hadronic stars (black) and hybrid stars with a
quark core (colored segments, stable part only). The inset enlarges the
region near the maximum mass and shows the unstable continuation of the
hybrid branches as dotted lines. Shaded boxes are the 68\% NICER
intervals for PSR~J0740$+$6620~\cite{Riley:2021pdl} and
PSR~J0030$+$0451~\cite{Riley:2019yda}.}
\label{fig:mr}
\end{figure*}

\begin{figure}[t]
\centering
\includegraphics[width=0.6\textwidth]{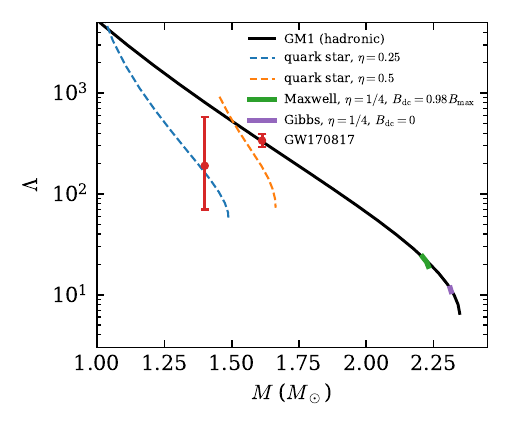}
\caption{Tidal deformability as a function of mass. Black: GM1 hadronic
stars. Dashed: compact branch of pure quark stars with flavor dressing.
Thick colored segments: stable hybrid stars. The error bar is the
GW170817 estimate for a $1.4\,M_\odot$ star at 90\%
credibility~\cite{LIGOScientific:2018cki}.}
\label{fig:lambda}
\end{figure}

The left panel of Fig.~\ref{fig:mr} shows the mass--radius relation of
stars made entirely of neutral CI quark matter, with flavor dressing, for
$1/4 \le \eta \le 3/2$. Every sequence displays two families of
configurations. At low central densities the star is a large object,
with radii between 20 and 60\,km, whose outer layers consist of the
dilute gas of constituent quarks found in Section~\ref{sec:quarkmatter}.
At high central densities the star contracts into a compact
configuration with a chirally restored core. For most values of $\eta$ the
two families are separated by an unstable segment on which
$dM/d\varepsilon_c < 0$, so that the sequence has two local maxima of the
mass. As $\eta$ grows, the maximum of the extended family rises faster
than that of the compact one, from $1.19$ to $2.62\,M_\odot$ against
$1.49$ to $2.09\,M_\odot$ with dressing, and for $\eta \ge 1$ the
largest masses are reached on the extended family.

\begin{table}[t]
\caption{Pure quark stars. Maximum mass and radius on the compact branch
($M_{\max}^{\rm c}$, $R^{\rm c}$) and on the extended branch
($M_{\max}^{\rm e}$, $R^{\rm e}$), and radius and tidal deformability of
a $1.4\,M_\odot$ star when it lies on the compact branch. Masses in
$M_\odot$, radii in km. A dash indicates that no separate compact branch
exists or that the $1.4\,M_\odot$ star belongs to the extended branch.}
\label{tab:QS}
\begin{tabular*}{\textwidth}{@{\extracolsep\fill}llcccccc}
\toprule
 & $\eta$ & $M_{\max}^{\rm c}$ & $R^{\rm c}$ & $M_{\max}^{\rm e}$ & $R^{\rm e}$ & $R_{1.4}$ & $\Lambda_{1.4}$ \\
\midrule
With dressing    & 1/4 & 1.49 & 12.0 & 1.19 & 34.1 & 13.5 & 162 \\
                 & 1/2 & 1.66 & 14.3 & 1.47 & 29.1 & -- & -- \\
                 & 3/4 & 1.86 & 17.1 & 1.78 & 26.2 & -- & -- \\
                 & 1   & 2.09 & 20.4 & 2.09 & 25.7 & -- & -- \\
                 & 3/2 & --   & --   & 2.62 & 26.6 & -- & -- \\
\midrule
Without dressing & 1/4 & 1.31 & 10.6 & 0.99 & 29.9 & -- & -- \\
                 & 1/2 & 1.47 & 13.0 & 1.28 & 25.3 & 15.2 & 308 \\
                 & 3/4 & 1.67 & 16.3 & 1.62 & 22.7 & -- & -- \\
                 & 1   & --   & --   & 1.93 & 22.4 & -- & -- \\
                 & 3/2 & --   & --   & 2.45 & 24.2 & -- & -- \\
\botrule
\end{tabular*}
\end{table}

Table~\ref{tab:QS} lists the relevant numbers. The compact branch never
reaches two solar masses with a radius compatible with the NICER
measurement of PSR~J0740$+$6620. For $\eta = 1/4$ with dressing it
produces a $1.4\,M_\odot$ star with $R_{1.4} = 13.5$\,km and
$\Lambda_{1.4} = 162$, both within the ranges allowed by NICER and
GW170817, but the maximum mass of the same sequence is only
$1.49\,M_\odot$. Increasing $\eta$ raises the maximum mass and
simultaneously moves the $1.4\,M_\odot$ configuration to the extended
family, where $R_{1.4}$ exceeds 30\,km and $\Lambda_{1.4}$ reaches
values between $10^4$ and $10^6$. Masses above $2\,M_\odot$ appear only
for $\eta \ge 1$ with dressing and $\eta \ge 3/2$ without it, always on
the extended family, with radii of 24 to 27\,km and central densities
of about $2\,n_0$. Figure~\ref{fig:lambda} makes the
same point in terms of $\Lambda(M)$: the sequences that pass through the
GW170817 interval stop well short of $2M_\odot$.

None of the sequences satisfies the mass and tidal constraints
together, and the configurations that come closest owe their mass to an
envelope of constituent quarks at densities where nuclear matter is
known to be hadronic. We regard this as evidence against pure CI quark
stars rather than as a tension to be resolved by further tuning. The
natural next step is to replace the unphysical low-density phase by a
realistic hadronic EoS and to ask where, if anywhere, deconfinement
takes place inside the star.

%% ====================================================================
\section{Hadronic matter and the deconfinement transition}
\label{sec:transition}
%% ====================================================================

\subsection{Hadronic equation of state}
\label{subsec:hadrons}

For the hadronic phase we use a relativistic mean-field model of
nucleons interacting through the exchange of $\sigma$, $\omega$, and
$\rho$ mesons, with cubic and quartic self-interactions of the scalar
field~\cite{Glendenning:1991es}. The energy density of nuclear matter
with proton and neutron densities $n_p$ and $n_n$ reads
\begin{equation}
\varepsilon_H = \sum_{i=n,p}\varepsilon_{\rm FG}(k_{F,i},m^*)
+ \frac{(m_N-m^*)^2}{2c_\sigma} + U(m_N-m^*)
+ \frac{c_\omega}{2}\,n^2 + \frac{c_\rho}{8}\,(n_p-n_n)^2,
\label{eq:epsH}
\end{equation}
where $n = n_p + n_n$, $m^*$ is the effective nucleon mass,
$\varepsilon_{\rm FG}$ is the energy density of a free Fermi gas,
$c_i = (g_i/m_i)^2$, and
$U(x) = \tfrac{b}{3}\,m_N x^3 + \tfrac{c}{4}\,x^4$. Rather than taking
the couplings from the literature, we fixed them by requiring the GM1
saturation properties~\cite{Glendenning:1991es}: $n_0 = 0.153\,\text{fm}^{-3}$,
binding energy $16.3$\,MeV, incompressibility $K = 300$\,MeV,
$m^*/m_N = 0.70$, and symmetry energy $a_{\rm sym} = 32.5$\,MeV. The
fit returns $c_\sigma = 11.785\,\text{fm}^2$, $c_\omega = 7.148\,\text{fm}^2$,
$c_\rho = 4.410\,\text{fm}^2$, $b = 0.002948$, and $c = -0.001071$, in
agreement with the published GM1 parameter set. Nucleons, electrons,
and muons are kept in $\beta$ equilibrium with zero net charge. Below
$0.57\,n_0$ the core EoS is joined to the SLy crust~\cite{Douchin:2001sv}.

Hadronic GM1 stars reach $M_{\max} = 2.35\,M_\odot$ at $R = 11.9$\,km,
with a central density of $5.67\,n_0$, and a $1.4\,M_\odot$ star has
$R_{1.4} = 13.7$\,km and $\Lambda_{1.4} \approx 810$. The model is thus
comfortably compatible with the $2M_\odot$  constraint but, like
other stiff mean-field models, exceeds the GW170817 bound on
$\Lambda_{1.4}$. We keep GM1 because it provides the off-equilibrium
pressure $P_H(\mu_B,\mu_e)$ needed for a Gibbs construction, which
tabulated EoSs do not. As a check that our conclusions do not hinge on
this choice, we repeated the Maxwell analysis with the tabulated SLy4
and DD2 EoSs~\cite{Douchin:2001sv,Typel:2009sy} distributed with
LALSuite~\cite{lalsuite}; our integrator reproduces their maximum masses,
radii, and tidal deformabilities (Appendix~\ref{app:checks}).

\subsection{Maxwell and Gibbs constructions}
\label{subsec:MG}

Hadronic and quark matter share two conserved charges, baryon number
and electric charge. If the surface tension between the two phases is
large, each phase stays locally neutral, and the transition occurs at
the baryon chemical potential where the neutral pressures coincide,
$P_H(\mu_B) = P_Q(\mu_B)$. This Maxwell construction produces a jump
in the energy density at constant pressure. If the surface tension is
small, the phases can carry opposite charges and coexist in a mixed
phase that is neutral only globally~\cite{Glendenning:1992vb}. The Gibbs
conditions then read
\begin{equation}
P_H(\mu_B,\mu_e) = P_Q(\mu_B,\mu_e),
\qquad
(1-\chi)\,q_H + \chi\,q_Q = 0,
\label{eq:Gibbs}
\end{equation}
where $\chi$ is the volume fraction of quark matter and $q_H$, $q_Q$
are the charge densities of the two phases including the common lepton
background. The energy density and baryon density of the mixed phase
are the corresponding volume averages. Gibbs matter has no density jump,
and the mixed phase begins at a lower pressure than the Maxwell
transition. The true behavior lies between the two limits and depends
on a surface tension that is poorly known, so we present both.

\subsection{Deconfinement bag shift and its upper bound}
\label{subsec:Bdc}

The CI does not confine quarks at finite density, and its quark pressure
carries no information about the energy cost of deconfinement. Following
the idea of the vBag model~\cite{Klahn:2015mfa}, we allow for a constant
shift of the quark pressure,
\begin{equation}
P_Q \to P_Q + B_{\rm dc}, \qquad
\varepsilon_Q \to \varepsilon_Q - B_{\rm dc},
\label{eq:Bdc}
\end{equation}
with $B_{\rm dc} \ge 0$ favoring the quark phase. Applied without
restriction, this shift has an unacceptable consequence. At
$\mu_B \approx 930$\,MeV the quark pressure of Eq.~\eqref{eq:Pq} vanishes,
so any positive $B_{\rm dc}$ would make quark matter more stable than
nuclei. We therefore admit the deconfined phase only where it is
physically meaningful, namely where the light quarks have restored
chiral symmetry, $\mu_u \ge \mu_\chi$. This mirrors the assumption in
vBag that deconfinement and chiral restoration occur together.

The restriction has a second consequence that fixes the size of the
shift. At the baryon chemical potential $\mu_B^*$ where neutral quark
matter first satisfies $\mu_u = \mu_\chi$, the shifted quark pressure
must not exceed the hadronic one; otherwise the pressure of the
favored phase would jump upwards at fixed $\mu_B$, which thermodynamics
forbids. This yields
\begin{equation}
B_{\rm dc} \le B_{\max} \equiv P_H(\mu_B^*) - P_Q(\mu_B^*).
\label{eq:Bmax}
\end{equation}
Table~\ref{tab:Bmax} lists $B_{\max}$ together with the hadronic density
at $\mu_B^*$, which is the earliest possible onset of deconfinement for
a given $\eta$. The bound grows with $\eta$, because vector repulsion
lowers the quark pressure, but the earliest onset density grows as well
and never falls below $3.4\,n_0$.

\begin{table}[t]
\caption{Upper bound on the deconfinement bag shift, Eq.~\eqref{eq:Bmax},
in MeV\,fm$^{-3}$, and hadronic density $n_H^*/n_0$ at the chemical
potential $\mu_B^*$ where the light quarks restore chiral symmetry in
neutral quark matter.}
\label{tab:Bmax}
\begin{tabular*}{\textwidth}{@{\extracolsep\fill}lcccc}
\toprule
 & \multicolumn{2}{c}{With dressing} & \multicolumn{2}{c}{Without dressing} \\
\cmidrule{2-3}\cmidrule{4-5}
$\eta$ & $B_{\max}$ & $n_H^*/n_0$ & $B_{\max}$ & $n_H^*/n_0$ \\
\midrule
1/4 & 68.8  & 3.41 & 98.5  & 3.69 \\
1/2 & 108.3 & 4.08 & 147.8 & 4.50 \\
3/4 & 148.9 & 4.46 & 196.2 & 4.88 \\
1   & 189.4 & 4.79 & 246.5 & 5.28 \\
\botrule
\end{tabular*}
\end{table}

%% ====================================================================
\section{Hybrid stars}
\label{sec:hybrid}
%% ====================================================================

\subsection{Where deconfinement sets in}
\label{subsec:onset}

\begin{figure}[t]
\centering
\includegraphics[width=0.6\textwidth]{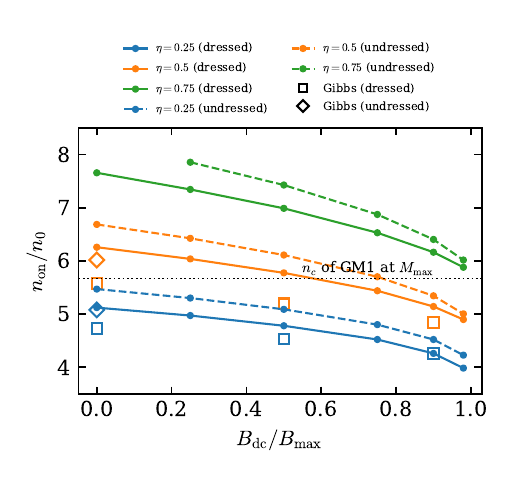}
\caption{Hadronic density at the onset of deconfinement versus the
deconfinement bag shift in units of its upper bound, Eq.~\eqref{eq:Bmax}.
Lines with dots: Maxwell construction with (solid) and without (dashed)
flavor dressing. Open symbols: Gibbs construction. The dotted horizontal
line is the central density of the maximum-mass GM1 star.}
\label{fig:onset}
\end{figure}

Figure~\ref{fig:onset} shows the hadronic density at which quark matter
first appears, as a function of $B_{\rm dc}/B_{\max}$, for the Maxwell
construction and, at selected values, for the Gibbs construction. The
dotted line marks the central density of the heaviest stable GM1 star,
$5.67\,n_0$. A quark phase can exist inside a stable star only if its
onset lies below this line.

Without any bag shift the Maxwell transition takes place at $5.1\,n_0$
for $\eta = 1/4$ and at $6.3\,n_0$ for $\eta = 1/2$ with flavor
dressing. The first value lies inside the GM1 sequence, but only
barely: the star whose center reaches the transition already has
$2.335\,M_\odot$, almost the maximum mass. For $\eta \ge 3/4$ the onset
lies at $7.7\,n_0$ or above, when it occurs at all below
$\mu_B = 2.4$\,GeV, and deconfinement never takes place in a stable
star. The tabulated EoSs lead to the same picture. With SLy4 and DD2 and
$B_{\rm dc}=0$, the Maxwell transition appears only for $\eta \le 1/2$,
at pressures between 360 and 820\,MeV\,fm$^{-3}$, and in every case at
or beyond the center of the maximum-mass configuration.

Raising $B_{\rm dc}$ lowers the onset monotonically. Near the upper
bound of Table~\ref{tab:Bmax} it reaches $4.0\,n_0$ for $\eta = 1/4$ and
$4.9\,n_0$ for $\eta = 1/2$ with dressing, and $4.2\,n_0$ and $5.0\,n_0$
without dressing. For $\eta = 3/4$ even $B_{\rm dc} = 0.98\,B_{\max}$
leaves the onset at $5.9\,n_0$ with dressing and $6.0\,n_0$ without
it, above the central density of the heaviest hadronic star. For
$\eta = 1$ the onset stays above $7\,n_0$. The bound of
Eq.~\eqref{eq:Bmax} is what prevents a further reduction: a larger shift
would pin the transition at $\mu_B^*$ with a discontinuous pressure.

The Gibbs construction moves the onset to lower densities, as expected.
For $\eta = 1/4$ with dressing and $B_{\rm dc} = 0$ the mixed phase
begins at $4.7\,n_0$ instead of $5.1\,n_0$ and extends up to about
$10\,n_0$, well beyond the stellar center. The gain shrinks as
$B_{\rm dc}$ increases. At $B_{\rm dc} = 0.9\,B_{\max}$ no mixed phase
forms at all, and the Gibbs result coincides with the Maxwell one. The
reason is the chiral condition of Section~\ref{subsec:Bdc}: in the mixed
phase the electron chemical potential is set by the hadronic component
and exceeds 200\,MeV, which pushes $\mu_u = \mu_B/3 - 2\mu_e/3$ below
$\mu_\chi$ and excludes the deconfined phase.

\subsection{Stellar sequences}
\label{subsec:hybridstars}

The right panel of Fig.~\ref{fig:mr} and Table~\ref{tab:hybrid} show the
stellar sequences for the cases in which the transition lies inside the
GM1 sequence. In all of them the quark core forms in stars between
2.21 and $2.35\,M_\odot$, and the stable hybrid branch is short. The
largest mass gain after the onset is $0.022\,M_\odot$, obtained with a
Maxwell transition for $\eta = 1/4$, flavor dressing, and
$B_{\rm dc} = 0.98\,B_{\max}$. The other cases gain less than
$0.01\,M_\odot$, and several of them become unstable as soon as the core
forms. We find no third family of compact stars in any of the cases
studied.

\begin{table}[t]
\caption{Hybrid stars with GM1 hadronic matter. Construction, vector
ratio $\eta$, bag shift, hadronic onset density, relative energy jump
$\Delta\varepsilon/\varepsilon_H$ (Maxwell), mass at the onset of the
quark core, and maximum mass. For Gibbs, that column gives instead the density
$n_{\rm end}/n_0$ at which the mixed phase ends. All cases with flavor dressing unless noted.}
\label{tab:hybrid}
\begin{tabular*}{\textwidth}{@{\extracolsep\fill}llccccc}
\toprule
 & $\eta$ & $B_{\rm dc}/B_{\max}$ & $n_{\rm on}/n_0$ &
 $\Delta\varepsilon/\varepsilon_H$ or $n_{\rm end}/n_0$ &
 $M_{\rm on}$ ($M_\odot$) & $M_{\max}$ ($M_\odot$) \\
\midrule
Maxwell & 1/4 & 0    & 5.12 & 0.94 & 2.335 & 2.335 \\
Maxwell & 1/4 & 0.90 & 4.26 & 0.26 & 2.257 & 2.260 \\
Maxwell & 1/4 & 0.98 & 3.98 & 0.12 & 2.209 & 2.230 \\
Maxwell & 1/2 & 0.98 & 4.90 & 0.22 & 2.323 & 2.323 \\
Gibbs   & 1/4 & 0    & 4.73 & 10.2 & 2.311 & 2.315 \\
Gibbs   & 1/4 & 0.50 & 4.53 & 8.8  & 2.294 & 2.298 \\
Gibbs   & 1/2 & 0.90 & 4.84 & 8.6  & 2.321 & 2.325 \\
Maxwell, no dressing & 1/4 & 0.98 & 4.23 & 0.15 & 2.253 & 2.262 \\
\botrule
\end{tabular*}
\end{table}

The size of the density jump explains why the branches are short. For
a Maxwell transition, small quark cores are stable when
$\Delta\varepsilon/\varepsilon_H < 1/2 + 3P_t/(2\varepsilon_H)$, where
$P_t$ is the transition pressure~\cite{Seidov:1971,Alford:2013aca}. All
the cases of Table~\ref{tab:hybrid} satisfy this condition, so the
hybrid branch is connected to the hadronic one. The limitation comes
from elsewhere. Because the onset lies so close to the central density
of the heaviest hadronic star, there is almost no room left before the
sequence turns over, and the soft quark EoS above the transition cannot
support much additional mass.

\subsection{Comparison with observations}
\label{subsec:obs}

The hybrid sequences inherit the maximum masses of GM1, between 2.23 and
$2.35\,M_\odot$, and the radii of their heaviest stars, between 11.9 and
12.7\,km, fall inside the NICER interval for PSR~J0740$+$6620. The
presence of a quark core changes the radius of the heaviest stars by at
most 0.12\,km relative to a purely hadronic star of the same mass, far below
current observational uncertainties.

Stars of intermediate mass are unaffected. A $1.4\,M_\odot$ star has no
quark core for any parameter set, so $R_{1.4}$ and $\Lambda_{1.4}$ are
those of GM1, as Fig.~\ref{fig:lambda} shows. The value
$\Lambda_{1.4} \approx 810$ exceeds the GW170817 bound, but this is a
property of the hadronic model and cannot be remedied by the quark
phase. A softer hadronic EoS lowers $\Lambda_{1.4}$, as SLy4 does
with $\Lambda_{1.4} = 297$, but it does not help the quark phase. With
SLy4 and $B_{\rm dc}=0$ the Maxwell transition moves up to
$P_t \approx 740$\,MeV\,fm$^{-3}$ for $\eta = 1/4$ and coincides with
the maximum-mass configuration, so the tidal constraint can be met only
by stars that contain no quark matter.

%% ====================================================================
\section{Conclusions}
\label{sec:conclusions}
%% ====================================================================

We have studied cold, neutral, $\beta$-equilibrated strange quark matter
with a contact interaction whose parameters are fixed entirely in
vacuum, and followed it into two kinds of compact stars. The three
flavors enter on the same footing, the vector repulsion generated by the
Fierz rearrangement is treated as a free fraction $\eta$ of the scalar
coupling, and the flavor dressing of the coupling is either switched on
or off to bracket its effect.

Regarding compact stars, those of pure quarks fail on two counts. The model does not produce
self-bound matter, so these objects carry an envelope of constituent
quarks that nature does not provide; and even setting that aside, no
value of $\eta$ yields both a $2M_\odot$ star and a tidal
deformability compatible with GW170817. The only sequences that reach
$2\,M_\odot$ do so on the extended branch, with radii near 25\,km.

Hybrid stars tell a sharper story. The vector strength that pure quark
stars need in order to become massive, $\eta \gtrsim 1$, is also the one
that pushes deconfinement beyond the center of the heaviest hadronic
star. For $\eta \lesssim 1/2$ the quark phase does appear, but never
below about $4\,n_0$, and the resulting cores live only in stars of 2.2
to $2.35\,M_\odot$ over a mass window of at most $0.02\,M_\odot$. This
holds for Maxwell and Gibbs constructions, with and without flavor
dressing, and for every admissible value of the deconfinement bag
shift. The upper bound on that shift, which follows from the continuity
of the pressure once deconfinement is tied to chiral restoration, is
what makes the statement robust: the one free parameter that could move
the transition is limited by thermodynamics rather than by choice.

These findings have a direct observational reading. Within this class
of models, the mass, radius, and tidal deformability of stars below
about $2.2\,M_\odot$ constrain only the hadronic EoS. Evidence for
deconfined matter would have to come from the heaviest pulsars, and
even there the imprint on the radius is about a tenth of a kilometer.
A softer hadronic EoS that satisfies the GW170817 bound does not
change this picture in the case we tested: with SLy4 the transition
moves to higher pressure and coincides with the maximum-mass star.

Several limitations should be kept in mind. We worked at zero
temperature and in the mean-field approximation, we did not include
hyperons in the hadronic phase, and we did not consider diquark pairing
in quark matter. Color superconductivity in particular adds a pairing
energy that raises the quark pressure at intermediate densities and
could lower the onset below the bound found here, in the same direction
as a positive $B_{\rm dc}$ but without the constraint of
Eq.~\eqref{eq:Bmax}; in NJL-type models, large pairing gaps are known to
stabilize quark cores in hybrid stars~\cite{Pagliara:2007ph}. Including the diquark channel of the contact
interaction, which is fixed by the same Fierz rearrangement and by
vacuum diquark masses, is therefore the natural extension of this work,
and it is the one that could overturn the conclusion that deconfinement
is confined to the heaviest neutron stars.

\backmatter

\bmhead{Acknowledgements}
A.\ Ahmad and N.\ Akhter acknowledge the hospitality of the Institute of
Physics, Gomal University. A.\ Raya acknowledges support from SECIHTI
(M\'exico) and CIC-UMSNH (M\'exico) under grants CBF-2026-465 and
18371, respectively.

%% ------------------------------------------------------------------
%% Declarations: TO BE COMPLETED AND AGREED BY ALL AUTHORS
%% The AI-use statement must describe the actual use (numerical
%% implementation, verification of calculations, and drafting), not
%% only language editing.
%% ------------------------------------------------------------------
%\section*{Declarations}
%Funding: see Acknowledgements.
%Competing interests: none.
%Data and code availability: [to be completed]
%Author contributions: [to be completed]
%Use of AI tools: [to be completed]

\begin{appendices}
%% ====================================================================
\section{Numerical checks}
\label{app:checks}
%% ====================================================================

The per-flavor gap equations, Eqs.~\eqref{eq:gapmed} and
\eqref{eq:mutilde}, are solved on a grid of quark chemical potentials
with a spacing of 1\,MeV up to $\mu_f = 950$\,MeV. At each point all
solutions are located by bracketing in $M_f$, and the one with the
lowest $\Omega_f$ is retained. The functions $\mathcal{C}$ and
$\mathcal{D}$ are evaluated in closed form and agree with direct
quadrature to machine precision. Thermodynamic consistency, tested by
comparing $P_q$ from Eq.~\eqref{eq:Pq} with the integral of $n_f$ over
$\mu_f$, holds to a relative accuracy of $10^{-5}$.

The TOV and tidal equations are integrated with an adaptive
eighth-order Runge--Kutta scheme, with relative tolerances between
$10^{-7}$ and $10^{-8}$. Density discontinuities are located as events
in the pressure, where the matching condition for $y$ is applied. We
tested the integrator against four cases with known results. For a bag
model with $\varepsilon = 3P + 4B$ and $B = 57\,\text{MeV\,fm}^{-3}$,
including the surface discontinuity, we find $M_{\max} = 2.015\,M_\odot$
at $R = 11.0$\,km. For an $n=1$ polytrope the Love number approaches
the Newtonian value $k_2 = (15-\pi^2)/(2\pi^2) \simeq 0.260$ as the
compactness decreases, reaching $k_2 = 0.256$ at $C = 2\times10^{-3}$.
For the SLy4 EoS~\cite{Douchin:2001sv} we obtain
$M_{\max} = 2.049\,M_\odot$, $R_{1.4} = 11.71$\,km, and
$\Lambda_{1.4} = 297$. For DD2~\cite{Typel:2009sy} with the unified
crust of the LALSuite tables~\cite{lalsuite} we obtain
$M_{\max} = 2.418\,M_\odot$, $R_{1.4} = 13.20$\,km, and
$\Lambda_{1.4} = 689$. These values agree with those reported in the
literature for the same EoSs.

For the hadronic model, the couplings obtained from the GM1 saturation
properties coincide with the published parameter set to the quoted
digits, and the saturation properties are recovered to better than one
part in $10^{8}$. In the Gibbs construction the hadronic field equations
are solved at fixed $(\mu_B,\mu_e)$ from several starting points, and
only converged solutions enter the search for the coexistence
condition. This precaution proved necessary: non-converged solutions
can produce spurious sign changes of the charge density and, with them,
false roots of the neutrality condition.

\end{appendices}

\bibliography{sn-bibliography}

\end{document}